\documentclass[12pt,a4paper]{article}
\usepackage{amssymb,amsmath,amscd,epsfig}
\title{Statistics of  neutrinos 
and the double beta decay}
\author{
A.S. Barabash$^{\rm a}$ , A.D. Dolgov$^{\rm a,b,c,d}$,  R. Dvornick\'y$^{\rm 
e}$,\\  
F. \v Simkovic$^{\rm e}$,  A.Yu. Smirnov$^{\rm d,f}$, 
\\[5mm]
${\rm ^a}$ {\small\it Institute of Theoretical and Experimental Physics, 
\small\it 117259 Moscow, Russia} \\
${\rm ^b}$\small\it{Dipartimento di Fisica, 
Universit\`a degli Studi di Ferrara, I-44100 Ferrara, Italy}
\\
${\rm ^c}$ {\small\it Istituto Nazionale di Fisica Nucleare,
Ferrara 44100, Italy} \\
${\rm ^d}$ {\small\it 
The Abdus Salam International Centre for Theoretical Physics, 
I-34100 Trieste, Italy}\\
${\rm ^e}$ {\small \it 
Comenius University, Dept. of Nuclear Physics and Biophysics, 
Mlynsk\'a dolina,}\\
{\small\it SK-84248 Bratislava, Slovakia}\\
${\rm ^f}$ {\small\it Institute for Nuclear Research, Russian Academy of
Sciences, Moscow, Russia} 
}
\date{}

\begin{document}

\newcommand{\beq}{\begin{equation}}
\newcommand{\eeq}{\end{equation}}
\newcommand{\be}{\begin{eqnarray}}
\newcommand{\ee}{\end{eqnarray}}
\newcommand{\bi}{\bibitem}
\newcommand{\lar}{\leftarrow}
\newcommand{\rar}{\rightarrow}
\newcommand{\lrar}{\leftrightarrow}
\newcommand{\mplq}{m_{Pl}^2}
\newcommand{\mnu}{m_\nu}
\newcommand{\nnu}{n_\nu}
\newcommand{\ngam}{n_\gamma}
\newcommand{\ms}{m_S}
\newcommand{\taus}{\tau_S}
\newcommand{\ns}{n_S} 
\newcommand{\me}{m_e}
\newcommand{\dnnu}{\Delta N_\nu}
\newcommand{\Tbbn}{T_{BBN}}
\newcommand{\nue}{\nu_e} 
\newcommand{\hf}{\hat f}
\newcommand{\hb}{\hat b}
\newcommand{\hfc}{\hat f^+}
\newcommand{\hbc}{\hat b^+}

\maketitle

\begin{abstract}
We assume that the Pauli exclusion principle is violated for neutrinos, 
and thus, neutrinos obey at least partly the Bose-Einstein statistics. 
The parameter $\sin^2 \chi$ is introduced that characterizes the 
bosonic (symmetric) 
fraction of the neutrino wave function. Consequences of the violation of 
the exclusion principle for the two-neutrino
double beta decays ($2\nu\beta\beta$-decays) are considered. This 
violation strongly changes the rates of the decays and modifies the energy 
and angular distributions of the emitted electrons. 
Pure bosonic neutrinos are excluded by the present data. In the case of partly 
bosonic (or mixed-statistics) neutrinos the analysis of the existing 
data allows 
to put the conservative upper bound $\sin^2 \chi < 0.6$. 
The sensitivity of future measurements of the 
$2\nu\beta\beta$-decay to $\sin^2 \chi$ is evaluated. 
\end{abstract}

\section{Introduction \label{s-intro}}

Does neutrino respect the exclusion principle of it's inventor?
In this paper we assume  that Pauli exclusion principle is 
violated for neutrinos and therefore neutrinos obey 
(at least partly) the Bose-Einstein  statistics.   
Possible violation of the exclusion principle was discussed 
in a series of  papers~\cite{ign-kuz} though no satisfactory 
and consistent mechanism of the violation  has been proposed so far. 
The assumption of violation of the Pauli exclusion principle 
leads to a number of fundamental problems which include   
loss of a positive definiteness of energy, violation of 
the CPT invariance, and possibly, of the  Lorentz invariance as well 
as of the unitarity of S-matrix.  
(For a critical review see ref.~\cite{lbo-rev}.) 
Experimental searches of the effects of the 
Pauli principle violation for electrons~\cite{exp-viol}  
and nucleons~\cite{exp-bar} have given negative results,  
leading to extremely strong bounds on the magnitude of 
violation.  
 
It may happen however that due to unique properties of neutrinos 
(neutrality, smallness of mass associated to some  
high mass scales), 
a violation of the Pauli principle in the  neutrino sector is  
much stronger than in other particle sectors. Therefore one 
may expect that  effects of its
violation can be first seen in neutrino physics.   

A possibility of the Bose statistics for neutrinos has been first 
considered in ref.~\cite{gri} where its effects on 
the Big Bang Nucleosynthesis (BBN) have been 
studied. According to \cite{gri} the change of neutrino 
statistics from pure fermionic to pure bosonic diminishes the 
primordial $^4{\rm He}$ abundance by $\sim 4\%$.   

The idea of bosonic neutrinos has been proposed independently 
in ref.~\cite{dosm}, where cosmological and astrophysical 
consequences of this hypothesis have been studied.  
Bosonic neutrinos might form a cosmological Bose condensate which 
could account for all (or a part of) the dark matter in the 
universe. ``Wrong'' statistics
of neutrinos modifies the BBN, leading to the
effective number of neutrino species smaller than three.
The conclusion in~\cite{dosm}  agrees qualitatively with  results of 
~\cite{gri} though quantitatively a smaller decrease of 
$N_{\nu}$ is found~\cite{hansen}. 

As far as the astrophysical consequences are concerned, 
dynamics of the  supernova collapse would be influenced and
spectra of the supernova neutrinos may change~\cite{dosm,kar}. 
The presence of neutrino condensate  would enhance contributions of the Z-bursts
to the flux of the UHE cosmic rays and lead to substantial
refraction effects for neutrinos from remote sources \cite{dosm}.\\

We assume that the Pauli principle is violated substantially 
for neutrinos, while the  violation is negligible for other particles. 
In particular, for electrons we will assume the usual Fermi-Dirac (FD) 
statistics. How to reconcile this pattern of the violation  
with the fact that in the standard model the left-handed
neutrino and electron belong to the same doublet?   
The answer may be connected to  the fact that neutrinos are the only
known neutral leptons and thus they can 
have substantially different properties from those of the charged 
leptons. In particular, neutrinos can be the  Majorana particles and 
violate lepton number conservation. 
The  difference between  charged leptons and neutrinos 
should be  related to  breaking of the electro-weak (EW) symmetry, and 
it can originate from  some high 
mass  scale of nature.  
One may consider  scenario where  violation of the Pauli 
principle occurs in  a hidden sector of  theory related to 
the Planck scale physics, or strings physics. 
It could be mediated by some singlets of the Standard model - (heavy) neutral 
fermions which mix with neutrinos when the EW symmetry is broken. 
Since only neutrinos can mix with the singlets, effects of the Pauli 
principle violation would show up first in the neutrino sector  and then 
communicate to other particles. 
In this way a small or partial violation of the relation between spin and 
statistics  might occur. 
A violation of the  spin-statistics theorem for other particles can be 
suppressed by  an additional power of a small parameter relevant for the 
violation in the neutrino sector and due to weak coupling of neutrino 
to other particle sector.\\

A violation of the Pauli principle for neutrinos
should show up in the elementary processes where identical
neutrinos are involved. A realistic process for  this test  
is the two-neutrino  double beta decay ($2\nu\beta\beta$-decay), 
\begin{equation}
A\rar A'+ 2\bar{\nu} + 2e^- 
\end{equation}
(or similar with neutrinos and positrons). 
It was shown in~\cite{dosm} that 
the  probability of the decay as well as the energy spectrum
and angular distribution of electrons should be affected. 
Qualitative conclusions were that the pure bosonic neutrino is excluded,  
whereas large fraction of the bosonic component 
in a neutrino state is still allowed by the present data. 
In this connection, a possibility of partly bosonic (mixed-statistics) 
neutrinos can be considered.\\

In this paper we perform 
a detailed study of the effects of bosonic 
neutrinos on the double beta decay. In sect. 2 we 
consider the  general case of partly bosonic neutrinos. 
We introduce a phenomenological parameter 
$\sin^2\chi$ which describes the fraction of bosonic 
neutrinos in such a way that a smooth change of 
$\sin^2\chi$ from 0 to  1 
transforms fermionic neutrinos into bosonic ones. 
So, in general,  neutrinos may  possess a kind of mixed or 
more general statistics than Bose or Fermi ones~\cite{para,ign-kuzm}.
In sect. 3 we present an analytic study of the double beta decay probabilities.
The exact expressions for the  $2\nu\beta\beta$-decay rates 
to ground and excited $0^+$  and
$2^+$ states with corresponding nuclear matrix elements (NME's) 
are given in sect. 4.
The results of numerical calculations of the total rates and various 
distributions for the $2\nu\beta\beta$-decays of $^{76}{\rm Ge}$ and 
$^{100}{\rm Mo}$ 
are presented in sect 5. In sect. 6. we obtain the bounds on  
$\sin^2 \chi$ from the existing data and  evaluate the sensitivities
of future double beta decay experiments. 
Discussion and conclusions are given in sect. 6.




\section{The $2\nu\beta\beta$-decay for  bosonic and partly bosonic neutrinos}

In the case of mixed statistics the operator of neutrino state can be 
written as 
\be
|\nu\rangle =  \hat a^+ |0\rangle  \equiv
c_\delta \hat f^+ |0\rangle + s_\delta \hat b^+ |0\rangle
= c_\delta | f\rangle + s_\delta | b\rangle
\label{nu-state}
\ee
where $| f\rangle$ and $| b\rangle$ are respectively 
one particle fermionic and bosonic states. 
The normalization of $|\nu \rangle$ implies $c^2_\delta + s^2_\delta =1$
($c_\delta \equiv \cos \delta$ and {$s_\delta \equiv \sin \delta$}).
$\hf$  ($\hf^+$) and $\hb$ ($\hb^+$) denote  fermionic, 
and bosonic annihilation (creation) operators.

To develop  a formalism for description of identical neutrinos 
one needs to specify  commutation/anti-commutation relations.
We assume that they have the following form:
\be
\hf \hb = e^{ i \phi} \hb \hf,\,\,\,
\hfc \hbc = e^{ i \phi} \hbc \hfc, \,\,\,
\hf \hbc = e^{ - i \phi} \hbc \hf,\,\,\,
\hfc \hb = e^{- i \phi} \hb \hfc,
\label{ab-af}
\ee
where $\phi$ is an arbitrary phase. Then the two-neutrino state can be  
defined as
\be
|k_1,k_2\rangle = \hat a_1^+ \hat a_2^+ |0\rangle. 
\label{two-nu}
\ee


For the pure bosonic neutrino one cannot introduce the Majorana mass term. 
So,  the neutrinoless double beta decay should be absent. In the case of 
partly bosonic neutrino, the neutrino mass would appear due to its fermionic 
component. This means that the kinematical mass measured,  {\it e.g.} in the 
tritium beta decay, would not be the same as the mass found from 
the neutrinoless 
beta decay. Such a situation, however, can be realized in the case of the usual 
fermionic neutrinos too.

The amplitude of the decay of nucleus $A \rightarrow 2\nu+2e+A'$ 
can be written as
\be
A_{2\beta} = 
\langle e(p_{e1}), e(p_{e2}), 
\overline\nu (p_{\nu 1}), \overline\nu (p_{\nu 2}),A'|
\int d^4 x_1 d^4 x_2
\psi_\nu (x_1) \psi_\nu (x_2) {\cal M}(x_1,x_2) 
| A \rangle.
\label{A-2beta}
\ee
After making the necessary commutation, according to  eq. (\ref{ab-af}), 
we obtain 
\be
A_{2\beta} = A_f \left[ c_\delta^4 + c_\delta^2 s_\delta^2
\left( 1-\cos \phi \right)\right]
+ A_b \left[ s_\delta^4 + c_\delta^2 s_\delta^2 \left( 1+\cos \phi \right)\right], 
\label{A-2beta-2}
\ee
where $A_f$ and $A_b$ are respectively fermionic (antisymmetric) 
and bosonic  (symmetric) parts of two antineutrino emission.
The amplitude can be parametrized as
\be
A_{2\beta} =\cos^2\chi\, A_f + \sin^2\chi\, A_b,
\label{A-2beta-3}
\ee
where $\cos^2\chi =  c_\delta^4 + c_\delta^2 s_\delta^2 \left( 1-\cos \phi \right)$
and  $\sin^2\chi =  s_\delta^4 + c_\delta^2 s_\delta^2 \left( 1+\cos \phi \right)$.

After integration over the neutrino phase space an interference between fermionic 
$A_f$ and bosonic $A_b$  parts of the amplitude $A_{2\beta}$ 
vanishes because the fermionic part is 
antisymmetric with respect to neutrino interchange, while bosonic is symmetric. 
The probability of the $2\nu\beta\beta$-decay is equal to:
\be
W_{tot} = \cos^4\chi\, W_f + \sin^4\chi\, W_b,  
\label{W-tot}
\ee
where $W_{f,b}$ are proportional to $|A_{f,b}|^2$. The expressions
for $W_{f,b}$ will be given in the next section.

Qualitative features of the $\beta\beta-$ decay in the presence of  the 
bosonic or partly bosonic neutrinos can be understood using the following 
consideration.  
Essentially, the effect of neutrino ``bosonization''  is that 
two contributions to the amplitude of the decay from diagrams with 
permuted neutrino momenta $p_{\nu 1} \leftrightarrow p_{\nu 2}$ 
should have relative plus sign instead of minus in the FD-case. 

The decay probability, $W_b$,  is proportional 
to  the bilinear combinations  of the type
$K^b_m K^b_n$,  $K^b_m L^b_n$, $L^b_m L^b_n$ (see the next section), where 
\begin{eqnarray}
K^b_m \equiv [E_m  - E_i + E_{e1} + E_{\nu 1}]^{-1} - [E_m  - E_i + E_{e2} + 
E_{\nu 2}]^{-1},
\nonumber\\
L^b_m \equiv [E_m  - E_i + E_{e2} + E_{\nu 1}]^{-1} - [E_m  - E_i + E_{e1} + 
E_{\nu2}]^{-1}. 
\label{prop}
\end{eqnarray}
Here $E_i$ is the energy of the initial nuclei, $E_m$ is the energy 
of the intermediate nuclei,  
$E_{ej}$, and $E_{\nu j}$ are the 
energies  of electrons and neutrinos respectively. 
The factors (\ref{prop}) correspond to the propagators of the 
intermediate nucleus. The key difference between
the bosonic and fermionic cases 
is the opposite signs of the two terms in the expressions (\ref{prop}).   
In the case of fermionic neutrinos 
they enter with the same signs (see, {\it e.g.}  \cite{boehm}):
\begin{eqnarray}
K^f_m \equiv [E_m  - E_i + E_{e1} + E_{\nu 1}]^{-1} + [E_m  - E_i + E_{e2} + 
E_{\nu 2}]^{-1},
\nonumber\\
L^f_m \equiv [E_m  - E_i + E_{e2} + E_{\nu 1}]^{-1} + [E_m  - E_i + E_{e1} + 
E_{\nu2}]^{-1}. 
\label{propf}
\end{eqnarray}
(Remember that for electrons we assume the normal 
Fermi statistics.) The terms in (\ref{prop}) correspond to 
the amplitudes with permuted momenta of both neutrinos and  
electrons. 
In the case of fermionic neutrinos such an interchange  
flips the sign twice (due to neutrinos and electrons), so that 
the overall sigh turns out to be plus.  
In the case of bosonic neutrinos the permutation of electrons only 
changes the sign, and the overall sign is minus. \\

Experimentally interesting are the 
$2\nu\beta\beta$-decays to  
the ground states  $0^+_{g.s.}$ and to excited states 
$0^+_1$ and $2^+_1$.  
The effect of 
bosonic neutrinos on the $2\nu\beta\beta$-decay half-life is different
for $J^\pi = 2^+$ and $J^\pi = 0^+$. This can be understood qualitatively, 
approximating the combinations  $K^b_m$ and $L^b_m$ for bosonic neutrinos  by 
\begin{equation}
K^b_m \approx \frac{E_{e2} - E_{e1} + E_{\nu 2} - E_{\nu 1}}{( E_m  - E_i + E_0/2 )^2}, 
~~~~L^b_m \approx \frac{E_{e1} - E_{e2} + E_{\nu 2} - E_{\nu 1}}{( E_m - E_i + E_0/2 )^2}, 
\label{KL}
\end{equation}
and the corresponding combinations  for the  fermionic neutrinos by  
\begin{equation}
K^f_m \approx L^f_m \approx  \frac{2}{E_m  - E_i + E_0/2 }.
\end{equation}
Here $E_0/2 \equiv \langle E_e + E_{\nu}\rangle $ is the average energy of 
the leptonic pair,  $E_0 \equiv E_i - E_f$ is the energy release in the decay, and $E_f$ 
is the energy of the final nucleus. 

For the $0^+ \rightarrow 0^+$ transitions 
an appearance of the differences of the electron 
and neutrino energies in the numerators of (\ref{KL}) 
leads to substantial (1-3 orders of magnitude) 
suppression of the total probability.  
It also modifies the energy distributions of electrons.

The effect of bosonic neutrinos on $0^+ \rightarrow 2^+$ 
transitions is opposite: The probabilities of transitions 
are proportional to the combinations 
$(K^{b}_m - L^{b}_m)(K^{b}_n - L^{b}_n)$,
where
\begin{equation}
(K^b_m - L^b_m) \approx  \frac{2(E_{e2} - E_{e1})}{(E_m  - E_i + 
E_0/2)^2}.
\end{equation}
In the case of fermionic neutrinos  
the combination $(K^f_m - L^f_m)$ has an additional factor 
$(E_{\nu2} - E_{\nu1})/(E_m  - E_i + E_0/2)$ and the suppression 
is stronger.  
Parametrically the probabilities of the $0^+ \rightarrow 2^+$ and 
$0^+ \rightarrow 0^+$ transitions become of the same order for  
bosonic neutrinos.

In the decay rates,  the kinematical factors $K^{f,b}_m$ and 
$L^{f,b}_n$ are weighted with the corresponding 
nuclear matrix elements (NME's).  
Let us introduce the ratio 
\begin{equation}
r_0 (J^\pi) \equiv \frac{W_b (J^\pi)}{W_f (J^\pi)},
\label{ratiow}
\end{equation}
of the decay probabilities to ground ($J^\pi= 0^+_{g.s.}$) 
and excited ($J^\pi= 0^+_{1},~ 2^+_{1}$) states
in pure bosonic $W_b(J^\pi)$  and pure fermionic
cases $W_f(J^\pi)$. In general, to find  $r_0(J^\pi)$ one needs to  
calculate the  NME 
for a given transition within an appropriate nuclear model.
The situation is simplified for those nuclear systems, where 
the transition via solely the ground state of the  intermediate 
nuclei $m=1$ dominates \cite{ABA84,SDS,DKSS}. 
For those nuclei the single state dominance 
(SSD) approximation (hypothesis) can be used.  
In this case the NME's can be factored out in  the rates  
and therefore  cancel in the ratio   $r_0(J^\pi)$. 

Let us consider the characteristics of the $\beta\beta$ decay to 
the ground and excited states $J^\pi$ in the mixed-statistic case 
of partly bosonic neutrinos. According to 
our considerations the total decay probability 
and the normalized total differential rate can be written as 
\begin{eqnarray}
W_{tot}(J^\pi) &=& \cos^4\chi W_f(J^\pi) + \sin^4\chi W_b(J^\pi),
\label{totpro}\\
\nonumber\\
P (J^\pi) &=& \frac{dW_{tot}(J^\pi)}{W_{tot}(J^\pi)} 
= \frac{\cos^4\chi\,
d\omega_f(J^\pi) + \sin^4\chi\, r_0(J^\pi) d\omega_b (J^\pi)}
{ \cos^4\chi + \sin^4\chi\, r_0(J^\pi)},   
\label{distr}
\end{eqnarray}
where 
\begin{equation}
d\omega_f(J^\pi) \equiv \frac{dW_f(J^\pi)}{W_f(J^\pi)}, 
~~~d\omega_b(J^\pi) \equiv \frac{dW_b(J^\pi)}{W_b(J^\pi)}
\label{normd}
\end{equation}
are the normalized distributions.  
Here $dW_f(J^\pi)$ and $dW_b(J^\pi)$  are the differential rates 
of the $2\nu\beta\beta$-decay for the pure fermionic and bosonic  
neutrinos. In the case of single state dominance due to factorization,  
the normalized  distributions do not depend on the uncertainties 
of the matrix elements \cite{SDS,DKSS}. 
In general, the factorization does not occur and the 
uncertainties of nuclear matrix elements restrict substantially the 
sensitivity of the $\beta\beta$-decay to statistics of neutrinos. \\

\section{Rates and nuclear matrix elements}

For the cases of pure fermionic and bosonic neutrinos we outline
the derivation of $2\nu\beta\beta$-decay rates. The relevant nuclear matrix
elements will be evaluated and discussed
using the  SSD and HSD (higher states dominance) hypothesis 
\cite{SDS,DKSS}. 

The matrix element of the $2\nu\beta\beta$-decay process takes the form
\begin{eqnarray}
<{f}|S^{{(2)}}|{i}> = 
\hspace{5cm}\nonumber \\
\frac{(-i)^2}{2} 
\int {<}e(p_{e1}), e(p_{e2}), 
\overline\nu (p_{\nu 1}), \overline\nu (p_{\nu 2}),A'|
T \left[ {\cal H}^\beta_{} (x_1) {\cal H}^\beta_{} (x_2)
\right] |A{>} dx_1 dx_2, 
\label{eq.7} 
\end{eqnarray}
where the weak $\beta$-decay Hamiltonian is  
\begin{equation}
{\cal H}^\beta_{} (x) = 
\frac{G_{F}}{\sqrt{2}} 
\left[\bar{e} (x)\gamma^\mu (1+\gamma_5) \nu_{e}(x)\right]
J_\mu(x) + {h.c.}.
\label{eq.8} 
\end{equation}
Here, $J_\mu(x)$ is the
weak charged (nuclear) hadron current in the Heisenberg 
representation. 
The $T$-product of the two hadron currents can be written as
\begin{eqnarray}
T(H^\beta_{} (x_{{1}}) H^\beta_{}(x_{{2}}))= 
\hspace{4cm}\nonumber \\
\Theta(x_{{10}} - x_{{20}})H^\beta_{}(x_{{1}})H^\beta_{}(x_{{2}}) + 
\Theta(x_{{20}} - x_{{10}})H^\beta_{}(x_{{2}}) H^\beta_{}(x_{{1}}). 
\end{eqnarray}

In the derivation of the $2\nu\beta\beta$-decay rate a number of 
conventional approximations have been used: i) Only the $s_{1/2}$ 
wave states of the outgoing leptons are taken into account. 
ii) The contribution of the double Fermi matrix element to the decay rate
is neglected as the initial and final nuclei belong to different
isospin multiplets. iii) Only the leading order 
($1/m_p$)  Gamow-Teller operators in the non-relativistic reduction
of the hadron current are retained. 

For the differential $2\nu\beta\beta$-decay rates 
to $0^+$ ground state and $2^+$ excited state we obtain 
\begin{eqnarray}
dW_{f,b}(J^+) = a_{2\nu} F(Z_f,E_{e1}) F(Z_f,E_{e2})
~{\cal M}^{f,b}_{J^\pi}~
d\Omega,   
\end{eqnarray}
where $a_{2\nu}=(G^4_\beta g_A)^4 m_e^9 /(64 \pi^7)$ 
and $G_\beta=G_F \cos\theta_c$ ($G_F$ is Fermi constant,
$\theta_c$ is Cabbibo angle). $F(Z_f,E_e)$ denotes the 
relativistic Coulomb factor and $g_A$ is the axial-vector
coupling constant. 
The upper index $f$ ($b$) stands for fermionic 
(bosonic) neutrinos. 

The phase space factor equals  
\begin{eqnarray}
d\Omega &=& \frac{1}{m^{11}_e}
E_{e1} p_{e1}~ E_{e2} p_{e2}~ E^2_{\nu 1}~ E^2_{\nu 2}
~\delta (E_{e1} + E_{e2} + E_{\nu 1} + E_{\nu 2} + E_{f} - E_{i})
\times \nonumber \\
&& ~~~~~~~~~~d E_{e1}~d E_{e2}~d E_{\nu 1}~d E_{\nu 2}~ d\cos\theta . 
\end{eqnarray}
Here, $\theta$ is the angle between the outgoing electrons. 
${\cal M}^{f,b}_{J^\pi}$ ($J^\pi = 0^+,~2^+$)
consists of the products of nuclear matrix elements: 
\begin{eqnarray}
{\cal M}^{f,b}_{0^+} &=&
\frac{m^2_e}{4} \left[ |{\cal K}^{f,b}_{0^+}+{\cal L}^{f,b}_{0^+}|^2 
+ \frac{1}{3}|{\cal K}^{f,b}_{0^+}-{\cal L}^{f,b}_{0^+}|^2 
\right] \nonumber\\
&&-\frac{m^2_e}{4} 
\left[ |{\cal K}^{f,b}_{0^+}+{\cal L}^{f,b}_{0^+}|^2 
- \frac{1}{9}|{\cal K}^{f,b}_{0^+}-{\cal L}^{f,b}_{0^+}|^2 \right]
~\frac{{\vec p}_{e1}\cdot {\vec p}_{e2}}{E_{e1}E_{e2}}, \nonumber\\
{\cal M}^{f,b}_{2^+} &=& m^2_e~
|{\cal K}^{f,b}_{2^+} - {\cal L}^{f,b}_{2^+}|^2 ~
\left(1+\frac{1}{3}
\frac{{\vec p}_{e1}\cdot {\vec p}_{e2}}{E_{e1}E_{e2}}\right)
\end{eqnarray}
with
\begin{eqnarray}
{\cal K}^{f,b}_{J^+} &=&
\frac{m_e}{\sqrt{s}}  
\sum_m <J^\pi_f||\sum_j \tau^+_j \sigma_j || 1^+_m>               
              <1^+_m||\sum_k \tau^+_k \sigma_k || 0^+_i> ~K^{f,b}_m \nonumber\\
{\cal L}^{f,b}_{J^+} &=&
\frac{m_e}{\sqrt{s}}  
 \sum_m <J^\pi_f||\sum_j \tau^+_j \sigma_j || 1^+_m>
               <1^+_m||\sum_k \tau^+_k \sigma_k || 0^+_i> ~L^{f,b}_m. 
\end{eqnarray}
Here, $s=1$ for $J=0$ and $s=3$ for $J=2$. $|0^+_i>$, $|0^+_f>$ ($|2^+_f>$)
and $|1^+_m>$ are, respectively, the states of the initial, final
and intermediate nuclei with corresponding energies $E_i$, $E_f$ and
$E_m$. The energy denominators  $K^{f,b}_m$ and $L^{f,b}_m$  
were introduced in Eqs. (\ref{prop}) and (\ref{propf}).

\subsection{Higher states dominance}
  
The $2\nu\beta\beta$-decay rates are usually evaluated in 
the  approximation in which the sum of the two lepton energies
in the denominator of the nuclear matrix element is replaced
with their average value $E_0/2$ 
\begin{equation}  
 E_m - E_i + E_{e j}+E_{\nu k} \approx E_m - E_i + E_0/2 
\end{equation}  
($j,k=1,2$). The main purpose of this approximation is to
factorize the lepton and nuclear parts in the calculation 
of the $2\nu\beta\beta$-decay half-life. This approximation
is justified if the transitions through the higher-lying 
states of the intermediate nucleus 
(at least few MeV above the ground state of (A,Z+1)
nucleus) give the dominant 
contribution to the $2\nu\beta\beta$-decay amplitude. This
assumption is called the higher states dominance (HSD)
hypothesis. It is expected to be realized for
A= 48, 76, 82, 130, 136 nuclear systems.

Assuming the HSD hypothesis we obtain for fermionic neutrinos  
\begin{eqnarray}
{\cal M}^{f}_{0^+} &\simeq& |M_{GT}^{(1)}(0^+)|^2
~\left(1-\frac{{\vec p}_{e1}\cdot {\vec p}_{e2}}{E_{e1}E_{e2}}\right), 
\nonumber\\
{\cal M}^{f}_{2^+} &=& |M_{GT}^{(3)}(2^+)|^2
\frac{(E_{e1}-E_{e2})^2~(E_{\nu 1}-E_{\nu 2})^2}{2 m^6_e}
\left(1+\frac{1}{3}
\frac{{\vec p}_{e1}\cdot {\vec p}_{e2}}{E_{e1}E_{e2}}\right).
\label{nmef}
\end{eqnarray}
In the case of bosonic neutrinos we end up with
\begin{eqnarray}
{\cal M}^{b}_{0^+} &=& |M_{GT}^{(2)}(0^+)|^2
~\left[ \frac{3(E_{\nu 2}-E_{\nu 1})^2+(E_{e 2}-E_{e 1})^2}{48 m_e^2}-
\right.\nonumber \\
&&~~~~~~~~~~~~~~~~~~
\left. \frac{9(E_{\nu 2}-E_{\nu 1})^2-(E_{e 2}-E_{e 1})^2}{144 m_e^2}
~\frac{{\vec p}_{e1}\cdot {\vec p}_{e2}}{E_{e1}E_{e2}}\right], 
\nonumber\\
{\cal M}^{b}_{2^+} &=& |M_{GT}^{(2)}(2^+)|^2
\frac{(E_{e1}-E_{e2})^2}{4 m^2_e}
\left(1+\frac{1}{3}
\frac{{\vec p}_{e1}\cdot {\vec p}_{e2}}{E_{e1}E_{e2}}\right).
\label{nmeb}
\end{eqnarray}
The Gamow-Teller matrix elements are given by 
\begin{equation}
M_{GT}^{(r)}(J^\pi) =
\frac{(2 m_e)^r}{\sqrt{s}}  
\sum_m \frac{<J^\pi_f||\sum_j \tau^+_j \sigma_j || 1^+_m>               
 <1^+_m||\sum_k \tau^+_k \sigma_k || 0^+_i>}
{(E_m~-~E_i~+~E_0/2)^r} 
\end{equation}
($r=1,2,3$). 

The full decay probabilities in pure bosonic $W_b$ and pure 
fermionic $W_f$ cases can be written as
\begin{eqnarray}
W_{f}(0^+) &=& |M_{GT}^{(1)}(0^+)|^2 {\cal I}^{f}_{HSD}(0^+), 
\nonumber\\
W_{f}(2^+) &=& |M_{GT}^{(3)}(2^+)|^2 {\cal I}^{f}_{HSD}(2^+) 
\end{eqnarray}
and
\begin{eqnarray}
W_{b}(J^\pi) = |M_{GT}^{(2)}(J^\pi)|^2 {\cal I}^{f}_{HSD}(J^\pi),  
\end{eqnarray}
where the phase space integrals are given by
\begin{eqnarray}
{\cal I}^{f,b}_{HSD}(J^\pi) = \frac{2 a_{2\nu}}{m^{11}_e}
\int_{m_e}^{E_i-E_f-m_e} 
f^{f,b}_{J^\pi}(E_{e1},E_{e2},E_{\nu 1},E_{\nu 2})
F_0(Z_f,E_{e1}) p_{e1} E_{e1} dE_{e1}\times
\nonumber\\ 
\int_{m_e}^{E_i-E_f-E_{e1}} F_0(Z_f,p_{e2}) p_{e2} E_{e2} dE_{e2}
\int_{0}^{E_i-E_f-E_{e1}-E_{e2}}  E_{\nu 2}^2  E_{\nu 1}^2  
d E_{\nu 1}
\end{eqnarray}
with $E_{\nu 2}= E_i -E_f -E_{e1}-E_{e2}-E_{\nu 1}$ and 
\begin{eqnarray}
f^{f}_{J^\pi}(E_{e1},E_{e2},E_{\nu 1},E_{\nu 2})
&=& 1~~~~~~~~~~~~~~~~~~~~
~~~~~~~~~~~~~~~~~~~~~~~ (J^\pi = 0^+),\nonumber\\
&=& \frac{(E_{e1}-E_{e2})^2~(E_{\nu 1}-E_{\nu 2})^2}{2 m^6_e}
~~~~~~~~~~ (J^\pi = 2^+),\nonumber\\
f^{b}_{J^\pi}(E_{e1},E_{e2},E_{\nu 1},E_{\nu 2})
&=& 
\frac{3(E_{\nu 2}-E_{\nu 1})^2+(E_{e 2}-E_{e 1})^2}{48 m_e^2}
~~~~~~ (J^\pi = 0^+),\nonumber\\
&=&
\frac{(E_{e1}-E_{e2})^2}{4 m^2_e}
~~~~~~~~~~~~~~~~~~~~~~~~~~~~ (J^\pi = 2^+).\nonumber\\
\end{eqnarray}
The $2\nu\beta\beta$-decay half-life is 
\begin{equation}
T^{f,b}_{1/2}(J^\pi) = \frac{\ln{2}}{W_{f,b}(J^\pi)}.
\end{equation}

\subsection{Single state dominance}

The single state dominance hypothesis assumes that the
$2\nu\beta\beta$-decays with $1^+$ ground state of the
intermediate nucleus (e.g., A=100, 116 and 128 nuclear systems)
are only governed by the two 
virtual $\beta$-transitions: i) the first one connects
the ground state of the initial nucleus with $1^+_1$
intermediate state; ii) the second one proceeds from
$1^+_1$ state to the final ground state. In this case
we find 
\begin{eqnarray}
{\cal M}^{f,b}_{0^+} &=& |M_{g.s.}(0^+)|^2
{m^2_e} \left[\frac{1}{3}(K^{f,b}K^{f,b}+{L}^{f,b}{L}^{f,b}+
{K}^{f,b}{L}^{f,b}) - \right.\nonumber \\
&&
\left. \frac{1}{9}(2 K^{f,b}K^{f,b}+2 {L}^{f,b}{L}^{f,b}+
5 {K}^{f,b}{L}^{f,b})
~\frac{{\vec p}_{e1}\cdot {\vec p}_{e2}}{E_{e1}E_{e2}}
\right], \nonumber\\
{\cal M}^{f,b}_{2^+} &=& m^2_e~|M_{g.s.}(2^+)|^2~
({K}^{f,b} - {L}^{f,b})^2 ~
\left(1+\frac{1}{3}
\frac{{\vec p}_{e1}\cdot {\vec p}_{e2}}{E_{e1}E_{e2}}\right)
\end{eqnarray}
with $K^{f,b}\equiv K^{f,b}_{m=1}$,
$L^{f,b}\equiv L^{f,b}_{m=1}$ and 
\begin{equation}
M_{g.s.}(J^\pi) =
\frac{1}{\sqrt{s}}  
<J^\pi_f||\sum_j \tau^+_j \sigma_j || 1^+_1>               
<1^+_1||\sum_k \tau^+_k \sigma_k || 0^+_i>.
\end{equation}

The value of the matrix element $M_{g.s.}(J^\pi)$
can be determined in a model independent way from the
single $\beta$-decay and electron capture measurements.
From the experimental values of 
$\log ~ft$ {\footnote {Because of wide range of $\beta$-lifetimes,
transitions are classified by $\log_{10} f t$ values (see e.g. \cite{behr}). 
$t$ and $f$ 
denote the measured half-life and the Fermi integral, respectively.}} 
for the electron
capture and the single $\beta$ decay of the ground state 
of the intermediate nucleus with $J^\pi= 1^+$ we obtain 
\begin{eqnarray}
|<1^+_1||\sum_k \tau^+_k \sigma_k || 0^+_i>|
&=& \frac{1}{g_A}\sqrt{\frac{3 D}{ft_{EC}}},
\nonumber\\
|<J^\pi_f||\sum_j \tau^+_j \sigma_j || 1^+_1>|               
&=& \frac{1}{g_A}\sqrt{\frac{3 D}{ft_{\beta^-}}}.
\end{eqnarray}
Here $D=G^4_\beta g^4_A/(8\pi^7)$.

Within the SSD approach for the full decay probabilities
we find
\begin{eqnarray}
W_{f,b}(J^\pi) &=& |M_{g.s.}(J^\pi)|^2 
{\cal I}^{f,b}_{SSD}(J^\pi),  
\label{ssdpi}
\end{eqnarray}
where 
\begin{eqnarray}
{\cal I}^{f,b}_{SSD}(J^\pi)
 = \frac{2 a_{2\nu}}{m^{11}_e}
\int_{m_e}^{E_i-E_f-m_e} 
g^{f,b}_{J^\pi}(E_{e1},E_{e2},E_{\nu 1},E_{\nu 2})
F_0(Z_f,E_{e1}) p_{e1} E_{e1} dE_{e1}\times
\nonumber\\ 
\int_{m_e}^{E_i-E_f-E_{e1}} F_0(Z_f,p_{e2}) p_{e2} E_{e2} dE_{e2}
\int_{0}^{E_i-E_f-E_{e1}-E_{e2}}  E_{\nu 2}^2  E_{\nu 1}^2  
d E_{\nu 1}
\end{eqnarray}
with
\begin{eqnarray}
g^{f,b}_{0^+}(E_{e1},E_{e2},E_{\nu 1},E_{\nu 2})
&=& 
{m^2_e} \left[\frac{1}{3}(K^{f,b}K^{f,b}+{L}^{f,b}{L}^{f,b}+
{K}^{f,b}{L}^{f,b})\right]\nonumber \\
g^{f,b}_{2^+}(E_{e1},E_{e2},E_{\nu 1},E_{\nu 2})
&=& 
{m^2_e} ~\left({K}^{f,b} - {L}^{f,b}\right)^2. 
\end{eqnarray}

\section{Characteristics of double beta decays}

In what follows we calculate the characteristics of the 
double beta decay mainly for two nuclei 
$^{100}{\rm Mo}$ and $^{76}{\rm Ge}$ for which the highest number of
events has been collected in experiment (see Ref. \cite{nemo}
and \cite{klapdor} respectively).

\subsection{Double beta decay of $^{100}{\rm Mo}$}

The NEMO-3 collaboration has detected about 
219 000  $(0^+ \rightarrow 0^+)$-decays of $^{100}{\rm Mo}$ 
\cite{nemo}. 
The signal to background ratio is very high  S/B =  
44 and the background is at the level of ~2.5\% only. 
All parameters of the decay: the sum of the electron energies, the energy of
each electron and the angular distribution (angular correlation of electrons) 
have been  measured. 

In the case of $^{100}{\rm Mo}$ the decay proceeds mainly 
through the $1^+$ intermediate nucleus and  
the single state dominance (SSD) hypothesis should give  a good 
approximation. This is also confirmed by spectra measurements in NEMO-3 
experiment~\cite{ARN04,SHI06}.  
Since $E_m - E_i \sim E_i - E_f $,  the lepton energies are important 
in the energy-denominators (\ref{prop}),  and consequently, in the rates.  

In the SSD approximation one can calculate the probability (NME)
using existing  experimental data for the beta-decay and the electron 
capture  of  $^{100}{\rm Tc}$  which is the intermediate dominating state. 
Accuracy of this  ``phenomenological''  calculation  is about  50\%, 
mainly because of poor experimental
accuracy for the electron capture process. 

Using the SSD approximation we calculated the 
$2\nu\beta\beta$-decay half-life of $^{100}{\rm Mo}$ to ground state  
for fermionic \cite{DKSS} and bosonic neutrinos (see sect. 3)  
\begin{equation} 
T_{1/2}^{f}(0^+_{g.s.}) = 6.8~10^{18} {\rm years}, ~~~
T_{1/2}^{b}(0^+_{g.s.}) = 8.9~10^{19}  {\rm years}, 
\end{equation}
so that the ratio of probabilities equals
\begin{equation}
r_0(0^+_{g.s.}) = 0.076.
\label{r0gs}
\end{equation}
The ratio $r_0(0^+_{g.s.})$ determines the weight with which  the bosonic
component  enters the total rate and differential distribution [see Eq.(\ref{totpro}]. 
For small $r_0$, a substantial modification of the distribution is expected for 
 $\sin^2 \chi$ being close to 1. 

The higher intermediate levels can give some (basically unknown) 
contribution  and this produces a systematic error in our analysis.  
To evaluate effect of the higher states, one can consider 
the extreme case described by the  higher states dominance (HSD) approximation,
which allows one to factorize the nuclear matrix element and 
integration over the phase space of outgoing leptons. In this case
the main contribution to the $2\nu\beta\beta$-decay matrix element comes
from the transition through higher energy states (including the region of 
the Gamow-Teller resonance)  of the intermediate nucleus.
Thus, the lepton energies in the denominators (\ref{prop}) can be neglected
(or approximated by $(E_f - E_i)/2$ ) due to a large value of $E_n - E_i$.
The fermionic and bosonic $2\nu\beta\beta$-decay rates are 
associated with different nuclear matrix elements [see Eq. (\ref{nmef})
and (\ref{nmeb})].
They can be evaluated within an appropriate nuclear model like
Quasiparticle Random Phase approximation (QRPA) or Nuclear Shell Model
(NSM). Then, the evaluated values of $2\nu\beta\beta$-decay half-life  and
ratio $r_0(0^+_{g.s.})$ are model dependent. Contrary, the normalized
differential characteristics are model independent for cases of pure fermionic
and bosonic neutrinos.

\begin{figure}[tb]
\begin{center}
\includegraphics[width=14.0cm, height=10.0cm, angle=0]{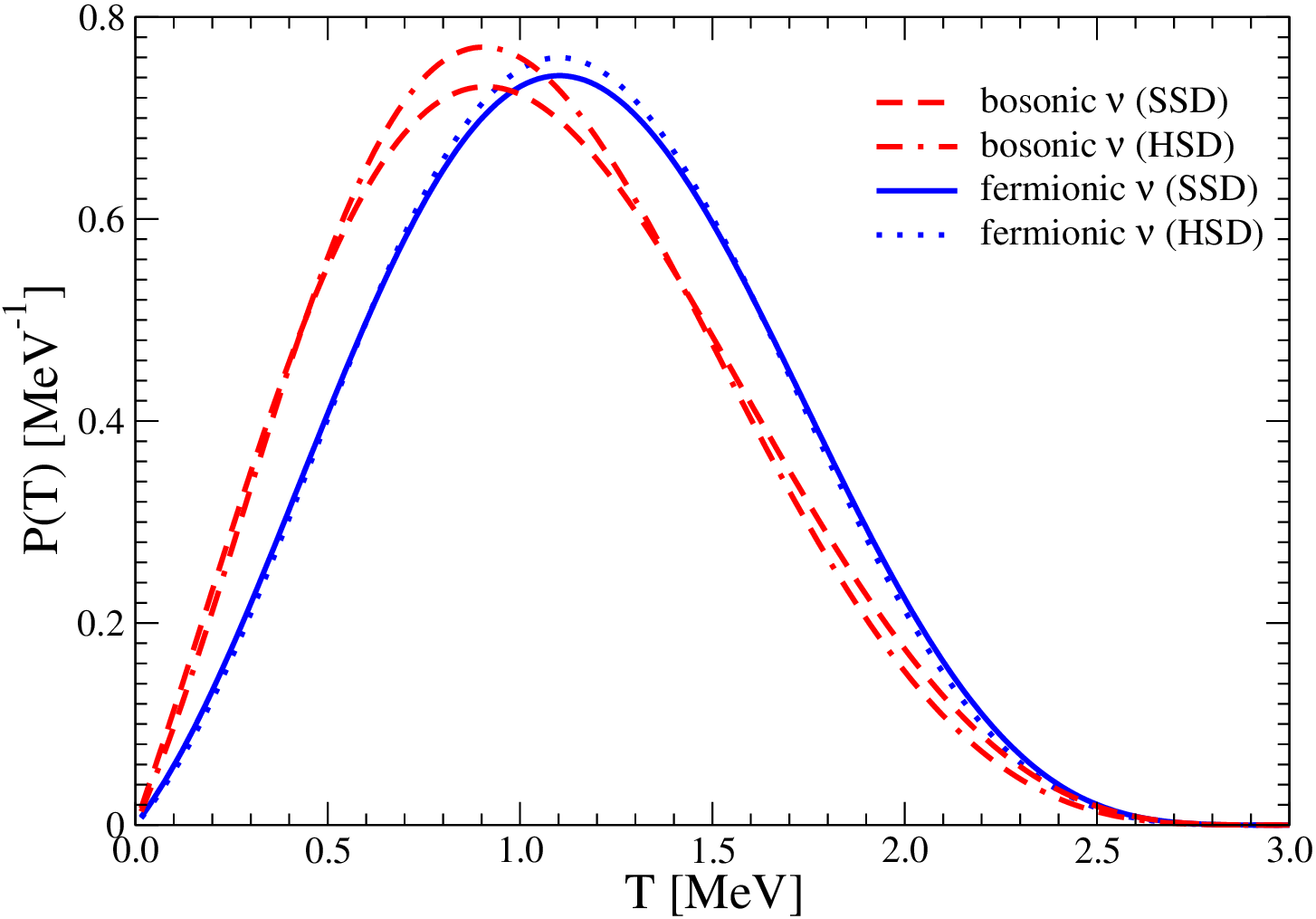}
\caption{The differential decay rates normalized to the total decay rate 
vs. the sum of the kinetic energy of outgoing electrons $T$ for 
$2\nu\beta\beta$-decay of $^{100}{\rm Mo}$ 
to the ground state of final nucleus.
The results are presented for the cases of pure fermionic and pure bosonic neutrinos.
The calculations have been performed within the single-state dominance hypothesis
(SSD) and with the assumption of dominance of higher lying states (HSD).
}
\label{mototapp}
\end{center}
\end{figure}

\begin{figure}[tb]
\begin{center}
\includegraphics[width=14.0cm, height=10.0cm, angle=0]{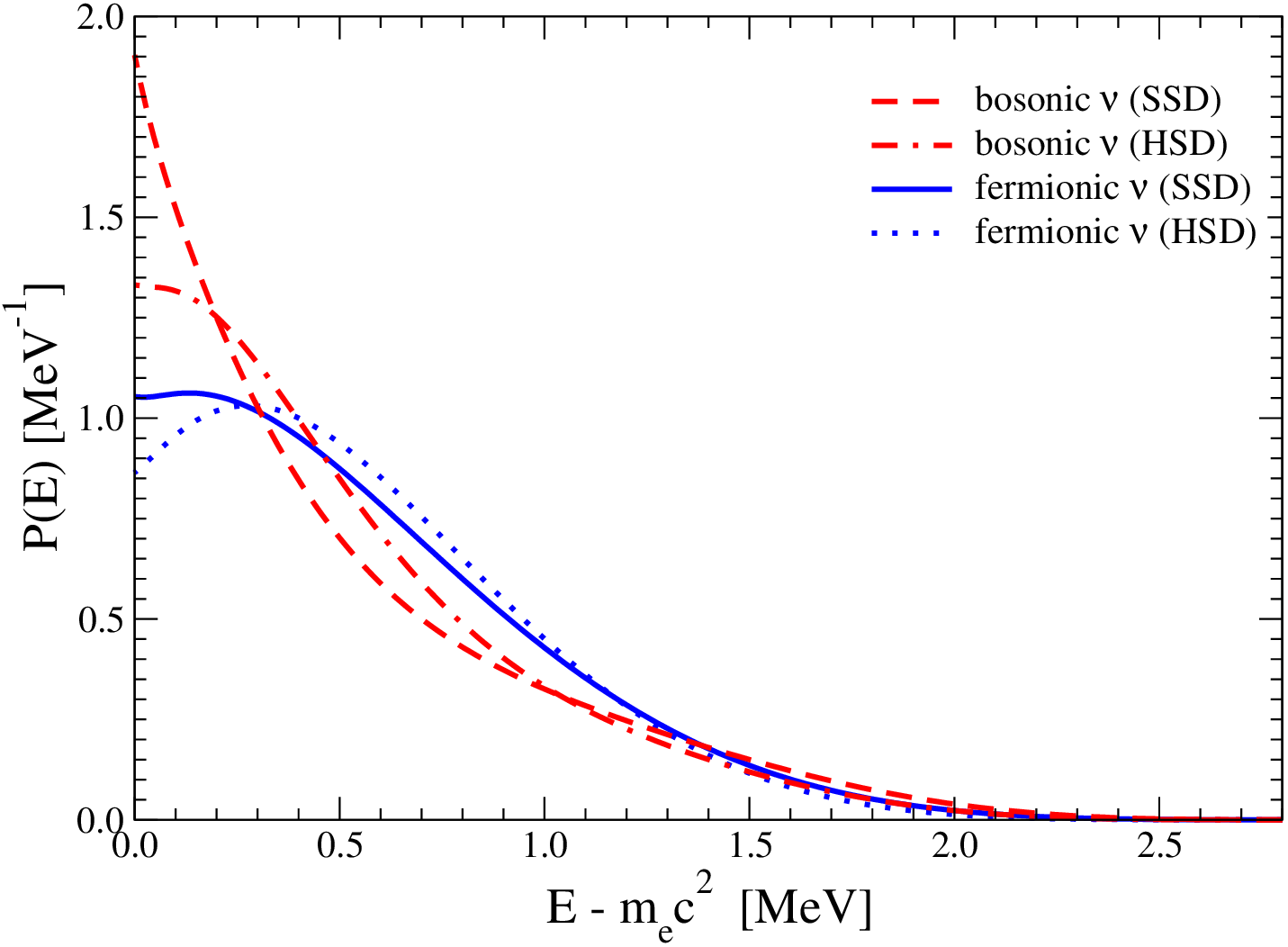}
\caption{The single electron differential decay rate normalized 
to the total decay rate vs. the electron energy for 
$2\nu\beta\beta$-decay of $^{100}{\rm Mo}$ 
to the ground state of final nucleus.
$E$ and $m_e$ represent the energy and mass of the electron, respectively.
The results are presented for the cases of pure fermionic and pure bosonic neutrinos.
The conventions are the same as in Fig. \protect\ref{mototapp}.
}
\label{mosinapp}
\end{center}
\end{figure}

The energy spectra of electrons calculated in the 
SSD and  HSD  approximations are presented 
in the figs. (\ref{mototapp}) and (\ref{mosinapp}). 
The SSD approximation gives slightly wider spectra of  
two electrons both for fermionic and bosonic neutrinos.  
The spectra for bosonic neutrinos are softer in both approximations. 
In particular, the  maxima  of SSD and HSD spectra are shifted to 
low energies for bosonic neutrinos by about 15 \% with respect 
to fermionic-neutrino  spectra. This shift does not 
depend on the approximation and therefore can be considered  as the solid 
signature of bosonic neutrino. 
Also the energy spectrum for single electron becomes softer in the bosonic 
case (Fig. \ref{mosinapp}).  

\begin{figure}[tb]
\begin{center}
\includegraphics[width=14.0cm, height=10.0cm, angle=0]{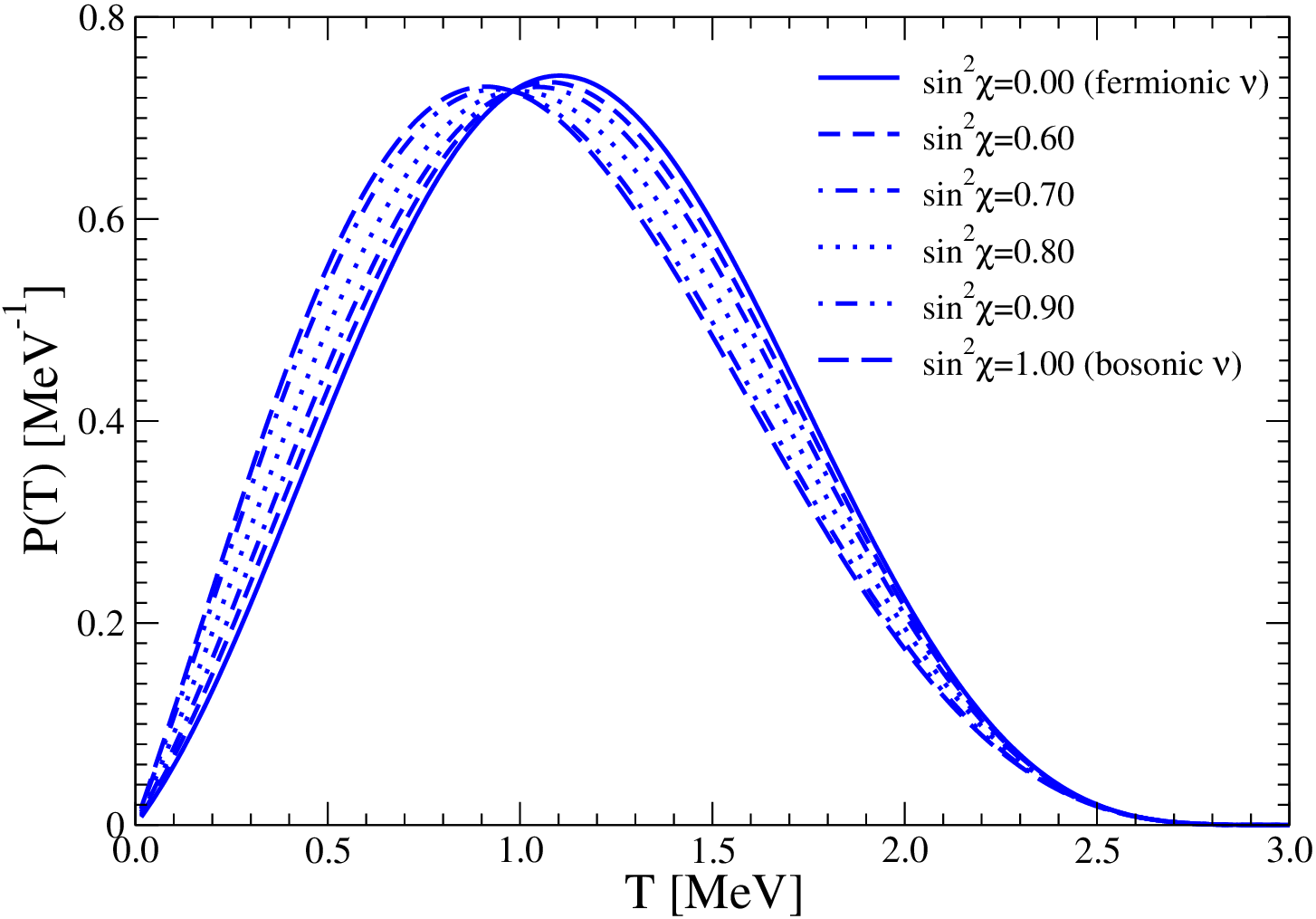}
\caption{The differential decay rates normalized to the total decay rate 
vs. the sum of the kinetic energy of outgoing electrons $T$ for 
$2\nu\beta\beta$-decay of $^{100}{\rm Mo}$ 
to the ground state of final nucleus.
The results are presented for different values of 
the squared admixture $\sin^2\chi$ of the bosonic component.
The spectra have been calculated in the SSD approximation.
} 
\label{mosum}
\end{center}
\end{figure}

In Fig.~\ref{mosum} we show the energy spectra of two electrons for 
different values of the bosonic-fraction $\sin^2 \chi$. 
With increase of  $\sin^2 \chi$ the spectra shift to smaller energies. 
Due to smallness of $r_0$ substantial shift occurs only when 
$\sin^2 \chi$ is close to 1.0 

\begin{figure}[tb]
\begin{center}
\includegraphics[width=14.0cm, height=10.cm, angle=0]{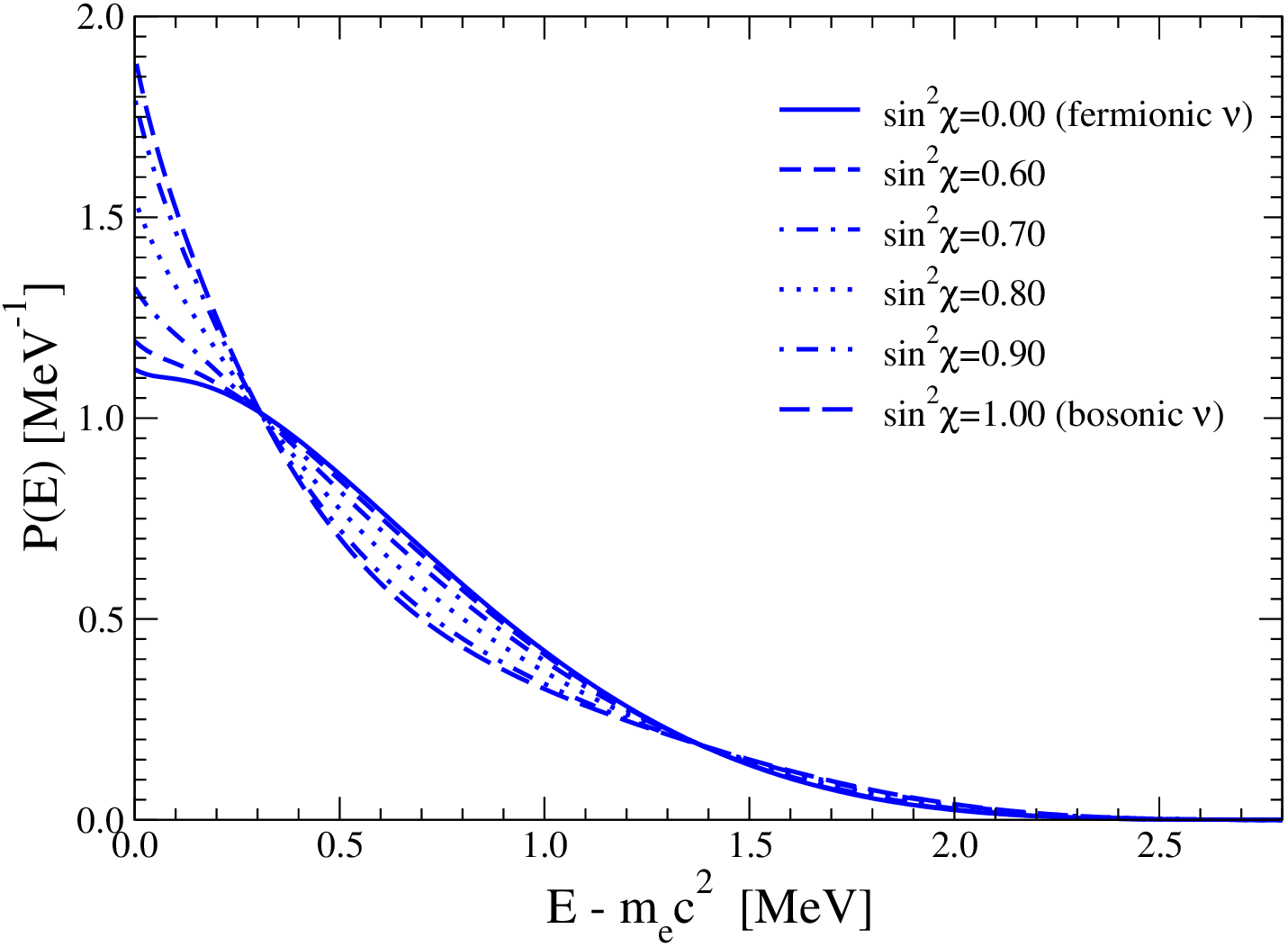}
\caption{
The single electron differential decay rate normalized 
to the total decay rate vs. the electron energy for 
$2\nu\beta\beta$-decay of $^{100}{\rm Mo}$ 
to the ground state of final nucleus.
The results are presented for different values of 
the squared admixture $\sin^2\chi$ of the bosonic component.
The spectra have been calculated in the
SSD approximation.
The conventions are the same as in Fig. \protect\ref{mosinapp}.
}
\label{mosingle}
\end{center}
\end{figure}

In Fig.~\ref{mosingle} we show the energy spectra of 
single electrons for different values of  $\sin^2 \chi$. 
A substantial change occurs at very low energies, with 
$E_{kin} = 0.3$ MeV being a fixed point.  
For $E_{kin} < 0.3$ MeV the distribution increase with  $\sin^2 \chi$, 
whereas for $E_{kin} =  0.3 - 1.4$ MeV it decreases.\\ 

As we mentioned before,  the rates of transitions to first excited
$2^+_1$ state are affected by the presence of bosonic neutrino component  
in the opposite (to $0^+$) way.  
Furthermore, in the SSD approximation the ratio of decay rates to the excited
$2^+$ state and to the $0^+_{g.s.}$ ground state does not depend on
the $\log ft_{EC}$ value, which is not measured accurately enough. 
For the $2\nu\beta\beta$-decay of $^{100}{\rm Mo}$ 
within the SSD approximation we obtain 
\be
T^{}_{1/2}(2^+_1) &=& 1.7~ 10^{23}~{\rm years} ~~~~~~~
({\rm fermionic}~\nu) \nonumber\\
&=& 2.4~ 10^{22}~{\rm years} ~~~~~~~
({\rm bosonic}~\nu).
\ee
Then the  ratio of the bosonic and fermionic half-lives equals
\be
r_0 (2^+_1) = 7.1.   
\ee
The bosonic rate is larger in agreement with our 
qualitative consideration in sect. 2.

The best lower bound on the $2\nu\beta\beta$-decay half-life to 
excited $2^+_1$ state is $1.6~10^{21}$ years \cite{BAR95}. 
The current limit of NEMO-3 experiment is $1.1~10^{21}$ years \cite{ARN07} 
(for 1 year of measurements). After 5 years of 
measurements with the present low-radon background conditions sensitivity will 
increase up to $\sim 10^{22}$ years thus approaching the 
prediction in the case of bosonic neutrinos. 
Due to the large value of $r_0$ even a small fraction of bosonic neutrinos 
can produce significant distortion of the standard (fermionic) 
spectra. 

\begin{figure}[tb]
\begin{center}
\includegraphics[width=14.0cm, height=10.0cm, angle=0]{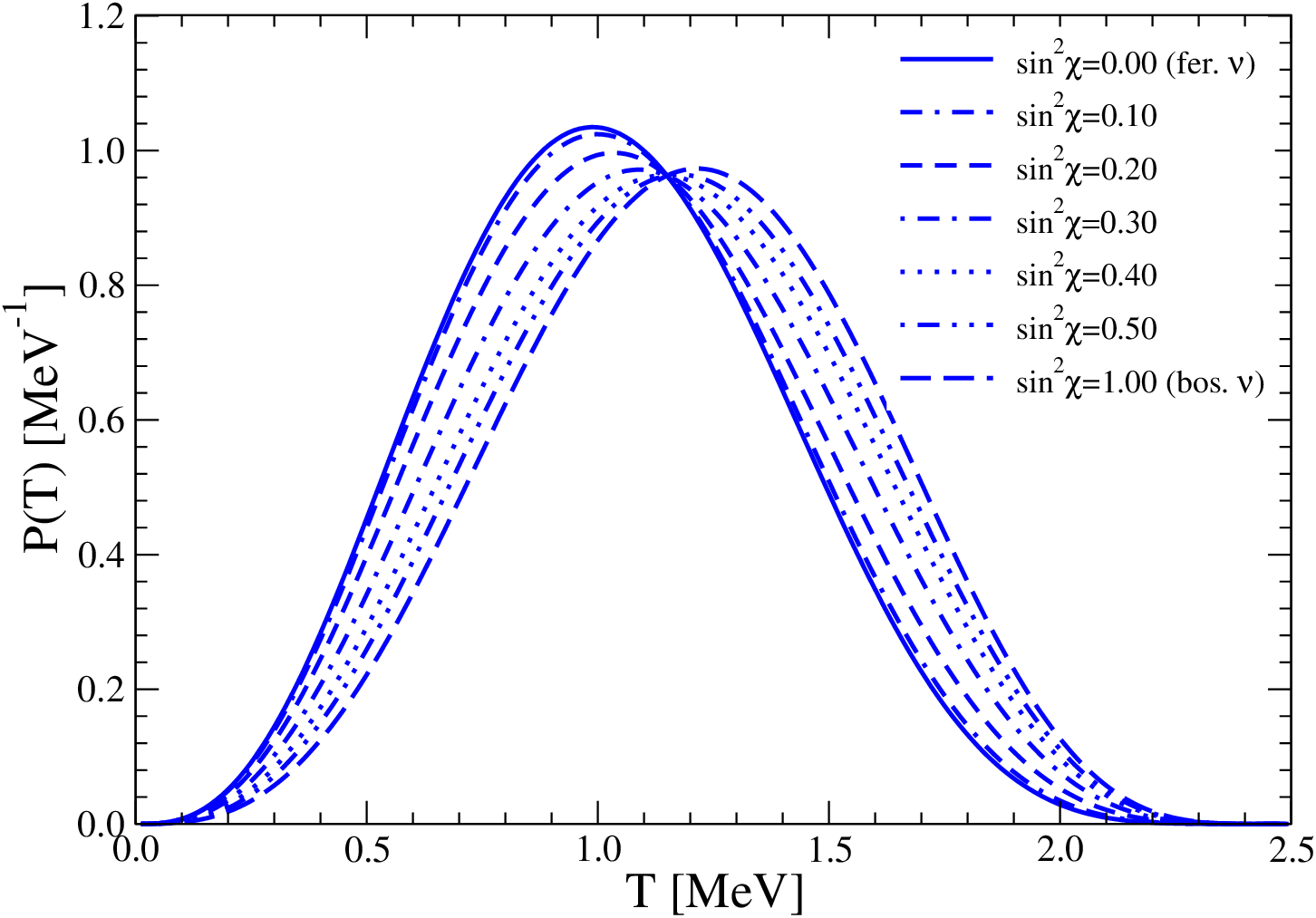}
\caption{
The differential decay rates normalized to the total decay rate 
vs. the sum of the kinetic energy of outgoing electrons $T$ for 
$2\nu\beta\beta$-decay of $^{100}{\rm Mo}$ to the excited $2^+_1$ state 
of final nucleus. 
The results are presented for different values of 
the squared admixture $\sin^2\chi$ of the bosonic component.
The spectra have been calculated in the
SSD approximation.
}
\label{mosumex}
\end{center}
\end{figure}

\begin{figure}[tb]
\begin{center}
\includegraphics[width=14.0cm, height=10.0cm, angle=0]{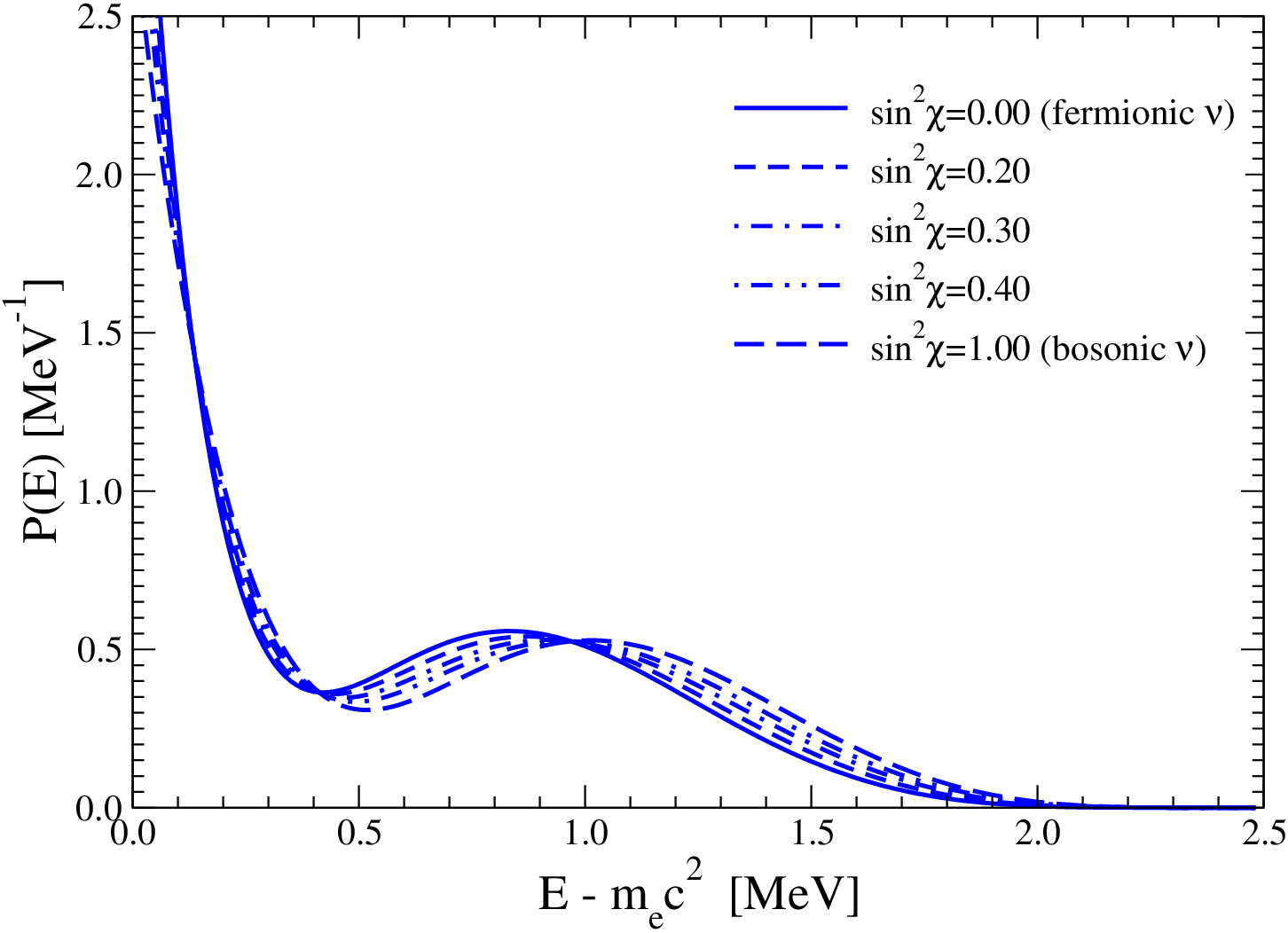}
\caption{
The single electron differential decay rate normalized 
to the total decay rate vs. the electron energy for 
$2\nu\beta\beta$-decay of $^{100}{\rm Mo}$ to the excited $2^+_1$ state 
of final nucleus.
The results are presented for different values of 
the squared admixture $\sin^2\chi$ of the bosonic component.
The spectra have been calculated in the SSD approximation.
The conventions are the same as in Fig. \protect\ref{mosinapp}.
}
\label{mosinex}
\end{center}
\end{figure}

Modifications of the spectra are opposite for the decay of  $^{100}{\rm Mo}$
into $2^+$ excited state: the spectra become harder with increase of 
$\sin^2\chi $ (see  Fig.~\ref{mosumex} and \ref{mosinex}). 
This is apparently related to the change of the spin of the nuclei. 
In the case of $0^+ -2^+_1$ transition the leptonic system should 
take spin 2 and therefore due to polarization of leptons 
(determined by V - A character 
of interactions) both electrons  move preferably in 
the same direction (hemisphere)  
and two antineutrinos in the opposite direction with the corresponding 
Pauli blocking factor. In the case of  bosonic neutrinos the Pauli blocking  
effect is reduced and therefore the electrons can be more aligned and consequently 
have higher energies. Correspondingly the spectrum becomes harder. 
In the case of $0^+ - 0^+$ transition the total leptonic momentum is zero, 
so that the electrons move in the opposite directions. 

According to Fig.~\ref{mosumex} even 10 $\%$ of "bosonic" admixture 
gives  substantial distortion effect and this fact can be used in the future 
experiments.   

The angular distribution of outgoing electrons \cite{SDS} can be 
written as 
\be
\frac{d W_{f,b}(J^\pi)}{d \cos \theta } =  
\frac{W_{f,b}(J^\pi)}{2} 
 (1 + \kappa^{f,b}(J^\pi) \cos \theta), 
\label{angular}
\ee
where $\theta$ is the angle between two electrons. 
For  $0^+ - 0^+$ transition and fermionic 
neutrinos in the SSD approximation  
\be
\kappa^f(0^+_{g.s.}) = -0.627 ~~~{\rm (fermionic~~ neutrino)}. 
\ee 
(The HSD approximation gives similar number: $ -0.646 $.)  
Notice that the preferable direction is $\theta = 180^{\circ}$ when electrons move in the 
opposite directions. The configuration with 
 the same direction of two electrons is suppressed.  
For bosonic neutrinos we find 
\begin{equation} 
\kappa^b (0^+_{g.s.}) = -0.344~~~   {\rm (bosonic~~ neutrino)}.  
\end{equation}
(The HSD approximation gives $-0.422$.)  
So,  the configuration with the same direction of electrons is less suppressed 
and the distribution is more isotropic (flatter) than in the fermionic case.

\subsection{$^{76}{\rm Ge}$ double beta decay}

The statistics of $^{76}{\rm Ge}$ decays
is  about  113000 events, the background is rather high, S/B =1.3,  and  only
the  sum of two electron energies is measured~\cite{klapdor}.
The  systematic error can be as large as 
10\%  and the main source of the error is the background. 
One has to estimate this background  independently and
make subtraction. 
So, one  can shift  the  spectrum
and its maximum within  the error. Furthermore, the energy spectrum of two
electrons  
starts to dominate over the background above  0.7 MeV  which means that 
the maximum of the spectrum is not observed. 
The advantage of $^{76}{\rm Ge}$ is that there is practically no difference 
between the results of HSD and SSD approximations for 
the energy distributions 
because the nearest $1^+_1$ state of the intermediate nucleus is 
lying high enough. Thus, one does not need to make assumptions 
about SSD or HSD.
In this way the conclusion does not depend on the nuclear structure details. 


In the  HSD approximation,  evaluating  the 
phase space integrals and nuclear matrix elements
within the proton-neutron QRPA we find 
\begin{equation} 
r_0 (0^+_{g.s.}) = 0.0014.
\end{equation}
This smallness is related to a large extend to  
high energies of the intermediate states, $E_m - E_i$ 
in comparison with leptonic energies restricted by the energy release 
$ E_l < (E_i - E_f)/2$: 
$E_l \ll E_m - E_i$. 
According to (\ref{prop}) the factors $K^b_m$, $L^b_m$ and consequently 
the rate are zero in the limit $ E_l = 0$. 
In the lowest approximation we obtain 
\begin{equation}
K^b_m, L^b_m \sim 
\frac{[(E_{\nu 2} - E_{\nu 1}) \pm (E_{e2} - E_{e1}) ]}{(E_m - E_i)^2}, 
\end{equation}
(where plus sign is for $K$-factors). 
Then the ratio of the rates can be estimated as 
\begin{equation}
r_0(0^+_{g.s.}) \sim \frac{\epsilon_l^2}{4(E_m - E_i)^2},  
\end{equation}
where $\epsilon_l$ is the  average energy of the lepton. 
Taking parameters of the  $^{76}{\rm Ge}$ -decay we find $r_0 \approx 10^{-3}$ 
in a good agreement with the calculations in QRPA.

\begin{figure}[tb]
\begin{center}
\includegraphics[width=14.0cm, height=10.0cm, angle=0]{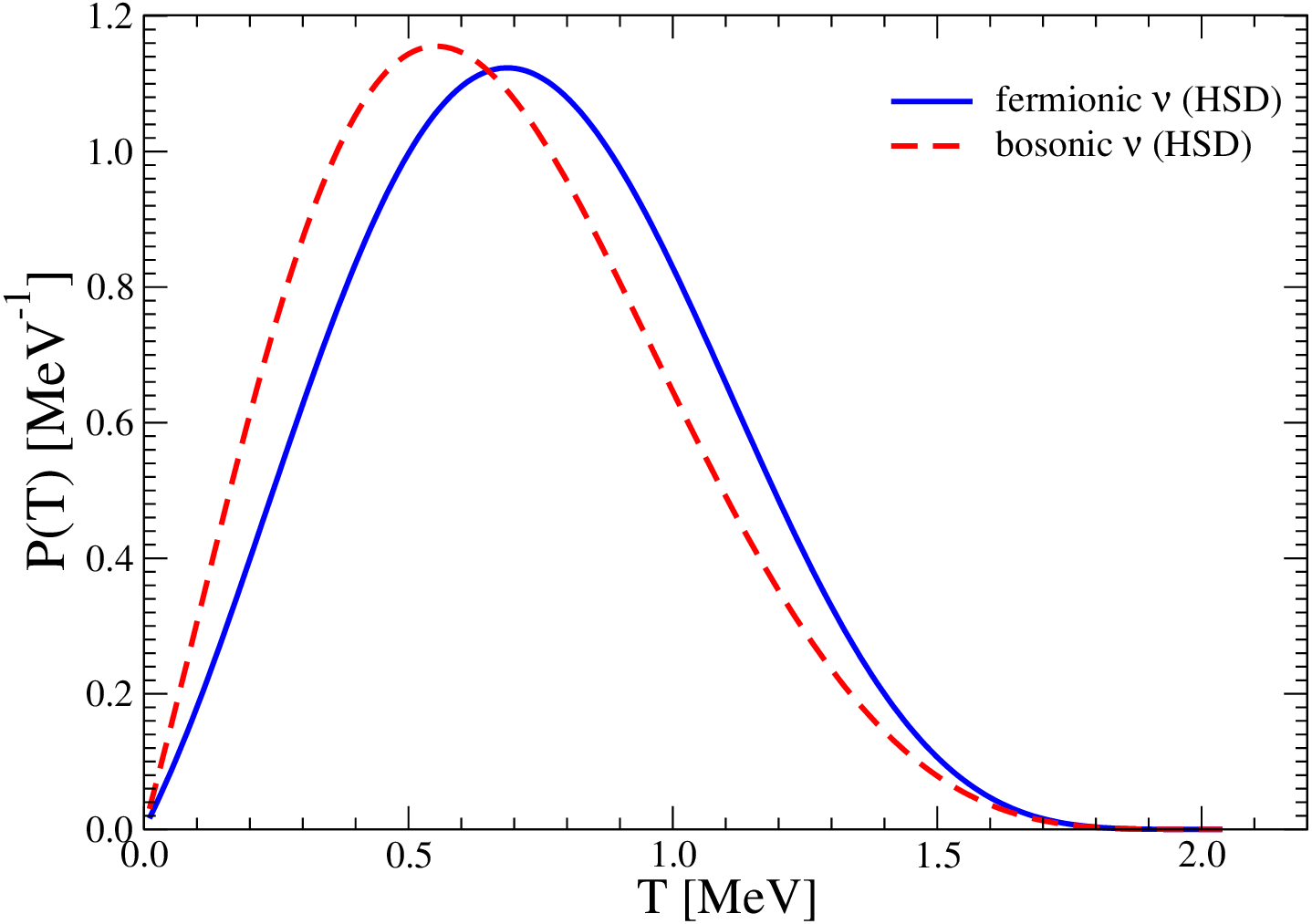}
\caption{
The differential decay rates normalized to the total decay rate 
vs. the sum of the kinetic energy of outgoing electrons $T$ for 
$2\nu\beta\beta$-decay of $^{76}{\rm Ge}$ to the ground state of final nucleus.
The results are presented for the cases of pure fermionic and pure bosonic neutrinos.
The calculations have been performed with the HSD assumption.
}
\label{gesum}
\end{center}
\end{figure}

\begin{figure}[tb]
\begin{center}
\includegraphics[width=14.0cm, height=10.0cm, angle=0]{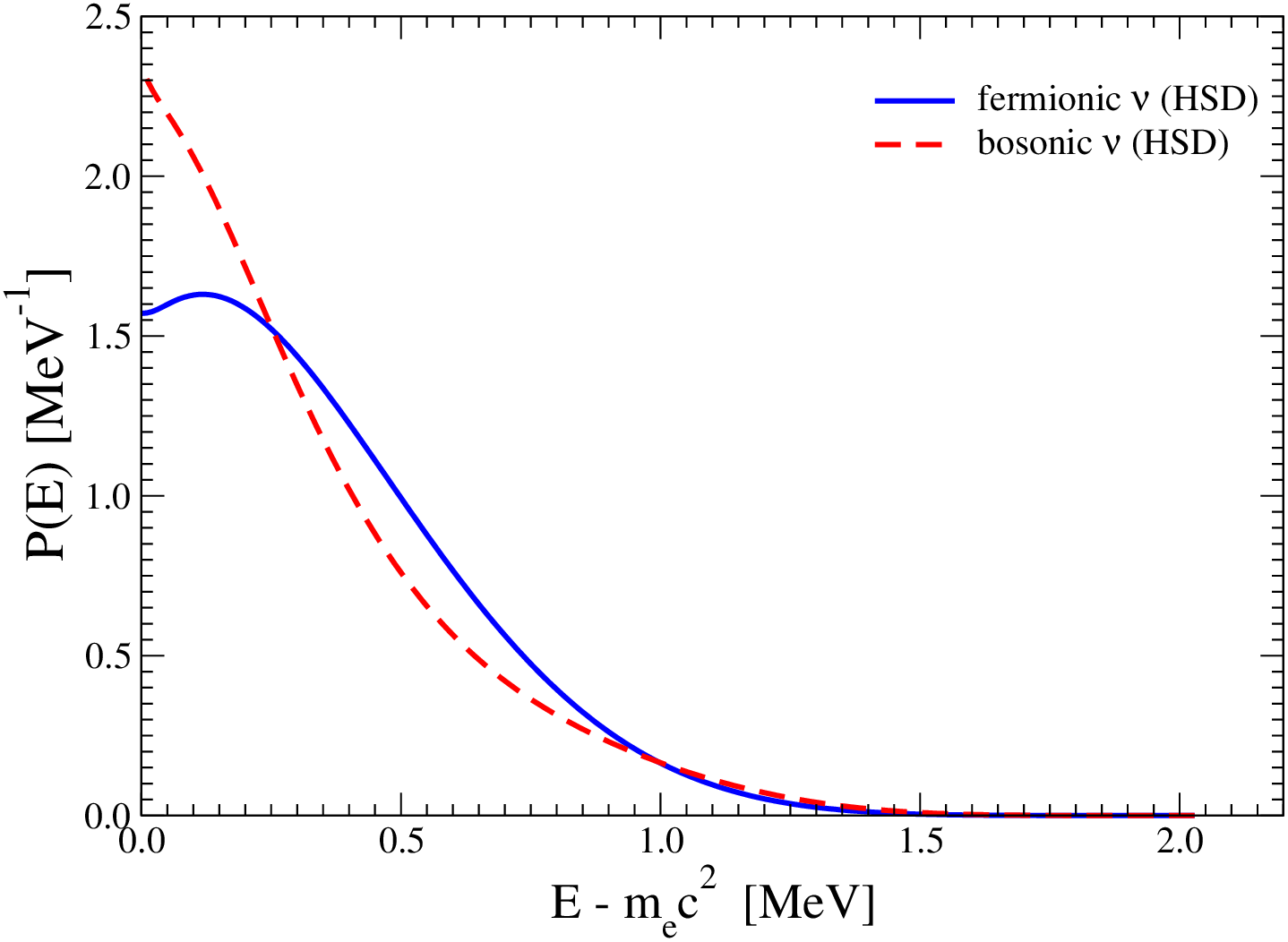}
\caption{
The single electron differential decay rate normalized 
to the total decay rate vs. the electron energy for 
$2\nu\beta\beta$-decay of $^{76}{\rm Ge}$ to the ground state of final nucleus.
$E$ and $m_e$ represent the energy and mass of the electron, respectively.
The results are presented for the cases of pure fermionic and 
pure bosonic neutrinos.
The calculations have been performed with the HSD assumption.
The conventions are the same as in Fig. \protect\ref{mosinapp}.
}
\label{gesin}
\end{center}
\end{figure}

In Fig.~\ref{gesum} we show the normalized distributions of the total 
energy of two electrons for pure fermionic and bosonic neutrinos. As in 
the case of $^{100}{\rm Mo}$, the  decay  with bosonic neutrinos 
has softer spectrum.  
The energy distribution of single electron is shown in Fig.~\ref{gesin}

Due to a small value of $r_0 (0^+_{g.s.})$ 
a substantial effect of the bosonic component 
should show up only for $\sin^2 \chi$ being very close to 1: 
$(1 - \sin^2 \chi)^2 \sim 10 r_0(0^+_{g.s.})$. 
So studies of the spectra are not sensitive to $\sin^2\chi$. 
In contrast, the total rate of the $^{76}{\rm Ge}$ decay gives 
a strong bound on $\sin^2\chi$. 

\section{Bounds on bosonic neutrinos}

One can  search for/restrict the 
bosonic  or partly bosonic neutrino
using total rates,  ratios of rates of the transitions to the 
excited and ground states, 
energy spectra, and  angular distributions.   
Let us evaluate the bounds on $\sin^2\chi$ that  can be obtained 
from the existing data using these methods.

As follows from our general discussion in sec.~3,  
for $0^+ \rightarrow 0^+$ transitions: $r_0 \ll 1$. 
For nuclei with small $r_0$ the best bound on bosonic neutrino fraction  
can be obtained from the total rates.  A modification of the spectrum 
due to presence of bosonic component is small. In contrast, the 
strongest modification of the spectrum  is expected 
for the nuclei with large $r_0$. 
This is true,   e.g., for $0^+ \rightarrow 2^+$ transition, 
where $r_0 \gg  1$.\\

{\it 1) Method 1:} Comparison of the predicted and measured  
half-life times. Using (\ref{totpro}) we can write 
\begin{equation}
\sin^2\chi = \frac{1}{1 + r_0}\left[1 - 
\sqrt{\frac{T^f_{1/2}}{T^{exp}_{1/2}} 
- r_0 \left(1 -\frac{T^f_{1/2}}{T^{exp}_{1/2}}\right)}
\right], 
\label{bound}
\end{equation}
where $r_0 = T^f_{1/2}/T^b_{1/2}$, 
$T^f_{1/2}$ ($T^b_{1/2}$)  are the theoretically predicted 
life-times for 
fermionic (bosonic) neutrinos and  $T^{exp}_{1/2}$ is the experimentally 
measured life-time. In the case of agreement between  the  measured  
and the predicted (for fermionic neutrinos) life-times, we can 
use (\ref{bound}) to establish the bound on parameter $\sin^2\chi$: 
\begin{equation}
\sin^2\chi < \frac{1}{1 + r_0}\left[1 - 
\sqrt{\frac{T^{f-min}_{1/2}}{T^{exp-max}_{1/2}} 
- r_0 \left(1 -\frac{T^{f-min}_{1/2}}{T^{exp-max}_{1/2}}\right)}
\right].
\label{bound1}
\end{equation}
Here, $T^{f-min}_{1/2}$ and $T^{exp-max}_{1/2}$ are, respectively, minimal theoretical 
value within a considered nuclear model (e.g., QRPA and its modification, NSM)
and maximal experimental value of the permitted experimental range
of the $2\nu\beta\beta$-decay half-life.
For  $r_0 \ll 1$ and  $r_0$ smaller the relative accuracy of determination of 
$T^f_{1/2}/T^{exp}_{1/2}$ the terms proportional to $r_0$   
in (\ref{bound1}) can be omitted. Then we get 
$\sin^2\chi < (1 - \sqrt{{T^{f-min}_{1/2}}/{T^{exp-max}_{1/2}}})$.

Apparently, this method  requires knowledge of the nuclear 
matrix element, and as 
we mentioned above,  reliable estimations  can be done for some nuclei 
e.g., $^{100}{\rm Mo}$ and $^{116}{\rm Cd}$  assuming SSD hypothesis. For some other 
nuclear systems nuclear models have to be considered. The two basic 
approaches used so far for the evaluation of the double beta decay 
matrix elements are the QRPA and the NSM. For the $2\nu\beta\beta$-decay 
of $^{76}{\rm Ge}$ the predicted half-lives 
are $7.7~10^{20}-1.4~10^{21}$ years (QRPA) \cite{SUH98} and 
$1.15~10^{21}$ years (NSM)~\cite{NSM}.
The experimental half-life (average half-life value is 
$(1.5\pm 0.1)~ 10^{21}$ years \cite{BAR06}) 
is in rather good agreement with the theoretical ones 
for fermionic neutrino within 
uncertainty characterized by the factor $\sim  2$  (see \cite{SUH98}).
For pure bosonic neutrinos $r_0(0^+_{g.s.}) \approx 10^{-3}$ 
(QRPA) and therefore 
for the  half-life time we would have 
$T^{b}_{1/2} \approx 1.5~ 10^{24} $ years, 
which is in contradiction with the experimental value. 
So, purely bosonic neutrino is certainly excluded. 
 
The axial-vector coupling constant $g_A$ is a significant source of 
uncertainty in the theoretical calculation of the $2\nu\beta\beta$-decay
rate, which is proportional to $g_A^4$. The commonly adopted values are 
$g_A=1.0$ (by assuming quenching in nuclear medium) and $g_A = 1.25$ 
(as for free nucleon). This gives about 1.5 uncertainty in NME's. 

For factor 2 uncertainty in NME we obtain factor 4 uncertainty 
in $T^f_{1/2}$. Therefore taking $T^f_{1/2} \sim T^{exp}_{1/2}$,  
we can put the  bound 
\begin{equation}
\frac{T^{f-min}_{1/2}}{T^{exp-max}_{1/2}} > \frac{1}{4}. 
\end{equation}
Then, eq. (\ref{bound1}) gives 
\begin{equation} 
\sin^2\chi <  0.50. 
\end{equation}
Notice that uncertainty in $T^f_{1/2}$ (and not $r_0$) dominates in this bound.

We can also use the half-life time of $^{100}{\rm Mo}$. 
Here $r_0(0^+_{g.s.})$ is much larger (\ref{r0gs})  but the accuracy of calculations of 
NME is better. 
Taking  SSD approximation we  can calculate the half life with 50\% accuracy:
$T^f_{1/2} = (6.84 \pm 3.42)~10^{18}$ years \cite{DKSS}. This value is in 
agreement with NEMO-3 value, 
$T^{exp}_{1/2} = (7.11 \pm 0.54)~ 10^{18}$ years \cite{nemo}. 
Plugging these numbers into (\ref{bound1}) we 
obtain for $r_0(0^+_{g.s.}) = 0.086$ 
\begin{equation}
\sin^2 \chi < 0.34.\,\,\,
\end{equation}
Notice that the accuracy of predicted half-life value is connected 
with experimental  accuracy for EC (electron capture)
half-life of $^{100}{\rm Tc}$ \cite{GAR93}. This accuracy can be 
improved in the future experiments{\footnote {In ref.~\cite{GAR93} 
Mo enriched to 97.4\% 
was used and the main background was 
connected with X-rays from different Tc isotopes which were produced in 
the sample due to (p,n) and (p,$\alpha$) reactions on different Mo isotopes, 
from $^{92}{\rm Mo}$ to $^{98}{\rm Mo}$; see Table II in~\cite{GAR93}. 
If one uses Mo enriched 
to 99\% (or more) then the mentioned above background would be much lower 
and the accuracy of the measurement would be several times better.}}
down to $\sim 10\%$ and 
correspondingly, the sensitivity to  $\sin^2\chi$
can reach   $\sim 0.1$. Unfortunately, there is only one (not very
precise) EC measurement for $^{100}{\rm Tc}$ and thus the above limit on 
$sin^2 \chi$ is not reliable enough.

Even stronger bound can be obtained from studies of $^{116}{\rm Cd}$ -decay. 
Recently a precise estimation of half-life 
value based on the  SSD approximation  and  information from the 
$^{116}{\rm Cd(p,n)}$ reaction was obtained: $T^f_{1/2} = (2.76\pm 0.12)~ 10^{19}$ years 
\cite{SAS07}. 
This prediction is in a very good agreement with experimental value 
(The experimental average is $(2.8 \pm 0.2)~ 10^{19}$ years 
\cite{BAR06}). Using these results we obtain from (\ref{bound1}) 
\begin{equation}
\sin^2 \chi < 0.06. \,\,\, 
\end{equation}
It should be noticed that the result of ref. \cite{SAS07} substantially differs 
from the earlier estimation $T^f_{1/2} =   (1.1 \pm 0.3)~ 10^{19}$ years
\cite{DKSS} (also based on SSD and measured value of electron capture rate
of $^{116}{\rm In}$ \cite{BHA98}). This result 
disagrees with the experimental value and could be interpreted 
as the effect of partly bosonic neutrino
with  $\sin^2 \chi   \sim 0.4$. \\



{\it 2) Method 2:}  Measurements of the differential characteristics of the decays: 
shapes of the energy spectra (sum energy 
and single
electron energy) and  angular distribution. Such information is provided  now 
by  NEMO-3 for
$^{100}{\rm Mo}$, $^{82}{\rm Se}$, 
$^{116}{\rm Cd}$, $^{150}{\rm Nd}$, $^{96}{\rm Zr}$ and $^{48}{\rm Ca}$. 
In the future the results for $^{130}{\rm Te}$ will be also available 
\cite{nemo,ARN04,SHI06,baranew}. 
In this method one 
compares the experimental  and theoretical energy spectra   
as well as the   angular distribution.
In practice one should perform the statistical fit of the spectra by  
a general distribution (\ref{distr}) with  $\sin^2 \chi$
being a free parameter.  
As we have seen the spectral method  has substantial sensitivity 
to $\sin^2\chi$ 
for nuclei and transitions with large $r_0$.  That includes  
$^{100}{\rm Mo}$, as well as transitions to the excited states. 
$^{76}{\rm Ge}$  with very small $r_0$ has no high sensitivity.

a) Let us consider first the energy spectra of  
$0^+_{g.s.} \rightarrow 0^+_{g.s.}$ decay of $^{100}{\rm Mo}$~\cite{nemo}. 
In the present paper we will not perform detailed statistical analysis 
of the spectra,  postponing this to the time when measurements  
will be finished and all
careful calibrations will be done. Instead, we give some 
qualitative estimates. There is a reasonable agreement between 
the predicted energy spectrum of  two 
electrons and  the experimental points. 
Therefore  we can certainly exclude the 
pure bosonic case ($\sin^2\chi  = 1$). 
Furthermore,  comparing the results of Fig.~\ref{mosum}
(essentially, the relative shift of the maximum of spectrum)
with the experimental spectrum  we can put the conservative bound  
$\sin^2 \chi < 0.6$.
In fact,  there is no ideal agreement between data and theoretical 
spectrum.  A better fit can be obtained  for $\sin^2\chi \sim 0.4-0.5$.  

b) Let us comment on the   
single-electron energy spectrum from $^{100}{\rm Mo}$ decay. 
The data reasonably well agree with the predictions from the fermionic 
SSD mechanism, but  
some difference exists between the data 
and the fermionic HSD-mechanism predictions. From this 
it was concluded that the  SSD mechanism is more relevant 
here \cite{ARN04,SHI06}. Comparing  the experimental 
data and spectra for partly bosonic neutrinos 
(Fig.~\ref{mosingle}) we  obtain:  $\sin^2\chi < 0.7$.  

Notice that the SSD spectrum does not show an
ideal agreement with data either. 
There is some discrepancy, especially in the low
energy region ($E = 0.2-0.4$ MeV). 
That could be explained by the effect of 
partly bosonic neutrinos with $\sin^2 \chi \sim$ 0.5 - 0.6.\\

Complete analysis 
of all existing NEMO-3 information (energy and angular distributions) 
using e.g. maximal likelihood methods, 
will have a higher sensitivity to $\sin^2 \chi$. However, 
it is difficult to expect a better bound than 
 $\sin^2 \chi \sim 0.4-0.5$, mainly because of the
existing disagreement between 
the data and Monte Carlo (MC) simulations. In fact,  
it can be just some systematic effect connected 
to the  present poor understanding of response function of 
the detector. If in future 
the NEMO experimental data turn out to be  in much better agreement with 
the MC-simulated spectrum, 
the  sensitivity to partly bosonic neutrino will be  improved 
down to $\sin^2 \chi= 0.2 - 0.3$.\\ 

{\it 3) Method 3:} Determination of the ratios of half-lives to excited and ground state, 
\begin{equation}
r^*_{f,b} (J^\pi) \equiv 
\frac{T^{f,b}_{1/2}(J^\pi)} 
{T^{f,b}_{1/2}(0^+_{g.s.})},
\label{ratio-ex}
\end{equation}
separately for fermionic and bosonic neutrinos.
For $2\nu\beta\beta$-decay of  $^{100}{\rm Mo}$ the ratio can be calculated 
rather reliably using 
the SSD-approximation.  The  advantage of this quantity
is that the EC amplitude, 
[(A,Z) $\rightarrow$ (A,Z+1) transition], which is not well determined, 
cancels in  the ratio (\ref{ratio-ex}). 

For $^{100}{\rm Mo}$  the transitions to the ground $0^+_{g.s.}$ 
and excited $0^+_1$ states 
were detected,  and in fact, some  discrepancy has been observed.  
The corresponding experimental ratio $r^*$ equals 
\begin{equation}
r^*_{exp.} (0^+_1) \simeq 80 
\end{equation}
(NEMO-3 results \cite{nemo,ARN07}),  
whereas within the SSD approach the  calculated ones are
\begin{eqnarray}
r^* (0^+_1) &\simeq& 61 ~~~~~~~({\rm fermionic}~\nu) \nonumber\\
&\simeq& 73 ~~~~~~~({\rm bosonic}~\nu).
\end{eqnarray}
A bosonic neutrino fits the data slightly better but the differences are 
probably beyond the accuracy of the SSD assumption. Still it is also 
necessary to  improve statistics in measurements of the transition to 
excited $0^+_1$ state.

Contrary to the case of $0^+$ excited state, the ratio 
of $2\nu\beta\beta$-decay 
half-lives to excited $2^+$ and ground state  is expected to be 
strongly different for bosonic and fermionic neutrinos. 
Using the SSD approximation for the 
$2\nu\beta\beta$-decay of $^{100}{\rm Mo}$ we found  
\begin{eqnarray}
r^* (2^+_1) &\simeq& 2.5~10^{4} ~~~~~~~({\rm fermionic}~\nu) \nonumber\\
&\simeq& 2.7~10^{2} ~~~~~~~({\rm bosonic}~\nu).
\end{eqnarray}
The $2\nu\beta\beta$-decay of $^{100}{\rm Mo}$ 
to excited $2^+_1$ state has been
not measured yet. Using the best experimental limit on the half-life
found in \cite{BAR95} we get
\begin{equation}
r^*_{exp} (2^+_1) > 2.2~10^{2}. 
\end{equation}
This bound is close to the bosonic prediction. A further experimental 
progress in measuring this nuclear transition will allow one to analyze also
the case of partially bosonic neutrino, and therefore is highly required.


\section{Conclusions}

\noindent
A study of the double beta decay  can provide a sensitive test of the Pauli 
exclusion principle and statistics of neutrinos. 
(Notice, that relation between the statistics of neutrinos and 
possible (small) violation of the Pauli principle is an open issue.) 
Appearance of the bosonic component in the neutrino 
states changes substantially 
the total rates of the decays as well as  the energy and angular distributions. 
We find, in particular,  that the ratio $r_0(0^+_{g.s.})$ of 
the rates to ground state 
for bosonic and fermionic neutrinos, is $< 10^{-3}$ 
for $^{76}{\rm Ge}$ and $0.076$ for $^{100}{\rm Mo}$, 
which excludes pure bosonic neutrinos. For transitions to $2^+$
excited states $r_0 (2^+) \gg 1$, in particular  
$r_0 (2^+_1)\simeq 7$. However, this 
$2\nu\beta\beta$-decay channel has been not measured yet.\\

\noindent
We have introduced phenomenological parameter 
$\sin^2\chi$ that describes the mixed 
statistics case of partly bosonic neutrinos. 
The dependence of the energy spectra and 
angular correlation of electrons on   $\sin^2\chi$ has been studied.
The bound on  $\sin^2 \chi$  can be obtained by comparison of the 
predicted and measured total rates of the decays. In spite of the 
big difference of the rates for fermionic and bosonic neutrinos, 
this method does not give strong and very reliable bound on  $\sin^2 \chi$
due to uncertainties  
in NME's.  The conservative upper bound  $\sin^2 \chi < 0.5$
 is found using  the $^{100}{\rm Mo}$ and 
$^{76}{\rm Ge}$ results. Much stronger bound,  
 $\sin^2 \chi< 0.06$,  is obtained  from recent studies of $^{116}{\rm Cd}$, 
however this bound requires further checks.\\ 

\noindent
The method based on the study of the normalized energy and angular spectra
is less affected by uncertainties in the NME's. 
The transitions with 
large $r_0(J^\pi)$ have the highest sensitivity to 
spectrum distortions and therefore  $\sin^2 \chi$. 
Using the data on 
the $0^+_{g.s.} \rightarrow 0^+_{g.s.}$ transition 
of  $^{100}{\rm Mo}$ we obtain the bound  $\sin^2 \chi<0.6$. 
In the future this bound can be improved down to  
$\sin^2 \chi\sim 0.2$.
The $0^+_{g.s.} \rightarrow 2^+_1$ transition with $r_0 (2^+_1) \simeq 7$ 
can give much stronger bound, but  here new, more sensitive experimental 
results are needed. 
We find that modification of the energy spectra due the presence of  
the bosonic components is opposite for $0^+_{g.s.} \rightarrow 0^+_{g.s.}$ 
and $0^+_{g.s.} \rightarrow 2^+_1$ transitions: 
for $0^+_{g.s.} \rightarrow 0^+_{g.s.}$  
the bosonic component leads to softer spectrum whereas for 
$0^+_{g.s.} \rightarrow 2^+_1$ transitions to harder spectrum of electrons. 
Also the presence of bosonic component 
leads to flatter angular ($\cos \theta$) 
distribution. \\

\noindent
Strong bound  (potentially down to  $\sin^2 \chi \sim 0.1-0.05$) 
might be obtained 
from measurements of ratios of the decay rates to the $2^+_1$ 
excited and ground state. However, this requires further experimental progress. \\

\noindent
We note that currently there are no restrictions on the admixture of bosonic component
from the BBN. However, as it was indicated in \cite{hansen} the future BBN studies 
will be able to constrain the fermi-bose parameter to $\kappa~ >~ 0.5$.    
The bound on parameter $\sin^2\chi~<~0.6$  from the $2\nu\beta\beta$-decay results in
$\kappa~ > - 0.2$.\\

\noindent 
In conclusion, the present data allow to put the conservative
upper bound on the admixture of the bosonic component 
 $\sin^2 \chi < 0.6$. With the 
presently operating experiments this bound might be improved down 
to $0.2$.  In future  one order of magnitude improvement seems feasible.


\section{Acknowledgments}

We are grateful to  L.B. Okun for helpful discussions.
F. \v S and A Yu. S. acknowledge the support of  the EU ILIAS project 
under the contract RII3-CT-2004-506222 and the VEGA Grant agency of the Slovak Republic 
under the contract No.~1/0249/03. 
A. Yu. S. is also grateful for support to the Alexander von Humboldt Foundation.  
This work was supported by Russian Federal Agency 
for Atomic Energy and by RFBR (grant 06-02-72553).

\end{document}